\documentstyle[epsfig]{article}

\begin{document}

\title{Space and Scale Localization in a Model of a Direct Transition to
Spatiotemporal Chaos}
\author{
Guoming Xiong$^*$\\
Department of Physics,
Lehigh University, Bethlehem, PA 18015\\
\\
J. D. Gunton\\
Department of Physics, 
Lehigh University, Bethlehem, PA 18015\\
\\
Haowen Xi\\
Department of Physics and Astronomy, \\
Bowling Green State University, Bowling Green OH 43403\\
\\
Ra\'ul Toral\\
Instituto Mediterr\'aneo de Estudios Avanzados (IMEDEA), CSIC-UIB\\
Campus Universitat de les Illes Balears, E-07071 Palma de Mallorca,
Spain}
\date{\today}
\maketitle
\begin{abstract}
We present the results of a wavelet-based approach to the study of the chaotic
dynamics of  a one dimensional model that shows a direct transition to spatiotemporal
chaos. We find that the dynamics of this model in the spatiotemporally chaotic regime
may be understood in terms of localized dynamics in both space and scale
(wavenumber). A projection onto a Daubechies basis yields a good separation of
scales, as shown by an examination of the contribution of different wavelet levels to
the power spectrum. At most scales, including the most energetic ones, we find
essentially Gaussian dynamics. We also show that removal of certain wavelet modes can
be made without altering the dynamics of the system as described by the Lyapunov
spectrum.

\end{abstract}
\vspace{1.5cm}


\section{Introduction}

A major goal in the study of temporal chaos in spatially extended systems
\cite{PM90,CH93,HG98} (spatiotemporal chaos or STC) is to find a statistical
description of the behavior of a particular dynamical system in the limit of
large length and long time scales. This is analogous to a hydrodynamic
description of a system of microscopic particles satisfying classical
mechanics.  Some progress has been made in this regard in recent years,
including constructing a long wavelength, long time theory of the
Kuramoto-Sivashinsky (KS) equation \cite{CH95}. In this reference, the authors
obtained an effective stochastic equation which belongs to the
Kardar-Parisi-Zhang (KPZ) university class in the hydrodynamic limit. This was
obtained by incorporating the chaotic dynamics of the small KS system in a
coarse-graining procedure. The basic premise of their approach is that the
spatiotemporal chaos of a large system can be understood in terms of the chaos
observed in mutually coupled, small systems. Two other recent studies
\cite{EBH96,WH99} of the Kuramoto-Sivashinsky equation used a wavelet
decomposition to characterize its spatiotemporal chaos. The first investigation
led to results similar to those of reference \cite{CH95}, suggesting a
statistical description of a group of identical short length subsystems, slowly
driven via interactions with the larger scales. More precisely, in \cite{EBH96}
it was shown that an effective equation could be obtained and consistently
approximated by a forced Burgers equation, for scales far from the cutoff
between small and large wavelengths. This work was extended in \cite{WH99}
where the authors found that projecting onto a spline wavelet basis enabled a
good separation of length scales, with each having its own characteristic
dynamics. At large scales they found essentially slow Gaussian dynamics, which
can be understood in terms of local events.  The results are also consistent
with the picture of weakly interacting small subsystems and so-called
"extensive chaos" (in which the Lyapunov dimension is proportional to $L^d$,
where $L$ is the linear system size and $d$ the dimension of space).  The
authors also discussed various correlation lengths and demonstrated the
existence of  a spatial interaction length, which provides a limit on how much
one may limit spatial interactions without changing the dynamics significantly
(and hence limits how small a system one can use).

Motivated by the success of the wavelet decomposition of the KS model, we have
undertaken such a decomposition for another one dimensional model, the so-called
Nikolaevskii equation \cite{BN93,TV96,TT96}. This model was originally proposed to
describe the propagation of longitudinal seismic waves. It was subsequently shown to
exhibit a direct transition from a spatially uniform, stationary state to a
spatiotemporally chaotic state as a control parameter was varied \cite{TT96}. The
spatiotemporal chaotic behavior of this system (which has been named as soft mode
turbulence or SMT) is different from that of the KS equation as a result of the
existence of an additional continuous symmetry in the model (beyond the conventional
symmetries of space and time translation invariance). This model is particularly
interesting in that there are two control parameters, $\epsilon$ and the system size
$L$, such that one can study the transition to chaos as a function of $\epsilon$ in
terms of power law behavior, scaling, etc.  In this sense it is a richer model than
the KS equation and many results have been obtained for it, including a calculation
of the Lyapunov exponents, Lyapunov dimension and Kolmogorov-Sinai entropy for
several values of $\epsilon$ in the limit of large $L$ such that extensive chaos
holds \cite{XTGT00}. It has also been shown that the distribution function for the
order parameter is Gaussian for large wavelengths and large times. In this paper we
carry out a wavelet decomposition similar to that in reference \cite{WH99},  using a
Daubechies basis. We find that the most energetic modes have a Gaussian distribution
for this choice of basis.  We also calculate the Lyapunov spectrum and find that one
can remove a certain set of the modes without altering this spectrum, suggesting that
in this sense one can obtain a more minimal description of spatiotemporal chaos for
this model.

The outline of the paper is as follows:  in section II we define the model and
the wavelet decomposition scheme.  In section III  we present the main results
of our analysis.  This includes a calculation of the energy distribution (power
spectrum) and the probability density functions for the wavelet
coefficients.  It also includes a calculation of the effects on the Lyapunov
spectrum resulting from removing various modes from the dynamics. In section IV
we present briefly the conclusions of this work.

\section{Model and Wavelet Decomposition}

The Nikolaevskii model is defined by the partial differential equation
\begin{equation}
\frac{\partial v}{\partial t} + \frac{\partial^{2}}{\partial x^{2}}
[\epsilon - (1 + \frac{\partial^{2}}{\partial x^{2}})^{2}]v +
v\frac{\partial v}{\partial x}= 0
\label{niko}
\end{equation}
with $0\leq x \leq L$ and periodic boundary conditions, where
$\epsilon$ and $L$ are two control parameters for
the model. In the present paper we study the model for
$\epsilon=0.5$ and $L\approx 158$; these values are large enough to see well
developed STC\cite{XTGT00}.\\
The general form of a discrete wavelet decomposition for a field
$v(x,t)$ can be written as
\begin{equation}
v(x,t)=\sum_{j}\sum_{k}a_{jk}(t)\Psi_{jk}(x)
\label{decom}
\end{equation}
where the set $\Psi_{jk}(x)=2^{j/2}\Psi(2^{j}x-k)$ form an
orthogonal basis in the sense that
\begin{equation}
\langle \Psi_{jk}\mid \Psi_{j'k'} \rangle \equiv
\int_{-\infty}^{\infty}\Psi_{jk}(x)\Psi_{j'k'}(x)dx=\delta_{j,j'}\delta_{k,k'}
\end{equation}
The indices $j,k$ are integers which
we specify below. The function $\Psi(x)$ is called the wavelet
function (or "mother" function) and the wavelet coefficients can be obtained as
\begin{equation}
a_{jk}(t)=\langle v \mid \Psi_{jk}
\rangle=\int_{-\infty}^{\infty}v(x,t)\Psi_{jk}(x)dx.
\label{coef}
\end{equation}
One obtains equations of motion for the wavelet coefficients by substituting
Eq.(\ref{decom}) into Eq.(\ref{niko}). Those equations of motion can then be
numerically solved in order to obtain the time evolution of the wavelet coefficients
and, hence, information about the chaotic behavior of the model. Alternatively, one
can solve directly Eq.(\ref{niko}) and use  Eq.(\ref{coef}) to obtain the wavelet
coefficients. This latter approach is the one we follow in this paper: at the required
times we decompose the resulting solution for $v(x,t)$ in terms of the Daub4
orthogonal basis set. The Daub4 is the simplest of a wavelet family named Daub$K$
constructed by Daubechies \cite{ID88} where $K$ ranges from $4$ to $20$. Based on this
wavelet family a very effective algorithm has been developed \cite{recipes}. Thus by
this method we obtain the time dependence of the wavelet coefficients $a_{jk}(t)$.

The Nikolaevskii equation (\ref{niko}) is integrated numerically using a version of
the pseudo-spectral method combined with a fourth-order predictor-corrector
integrator.  (The details of this method will be given elsewhere). We use a mesh size
$\delta x =0.31$ and a time step $\delta t=0.01$ in the numerical scheme.  The length
$L$ of the system is $L=N\delta x$, where $N$ is the total number of points of the
system. It is convenient to choose $N=2^{J+1}$ ($J$ is an integer which represents
the largest level of the wavelet) in order to apply Daub4 in an efficient pyramidal
scheme. In our case we usually choose $J=8$ so $N=512$ and $L=158.72$.  The total
integration time in our simulation goes from $t=600,000$ to $t=1,100,000$ (in the
dimensionless units of  equation (\ref{niko})), depending on which quantity we
calculate. Usually, once we have reached the chaotic state, we perform time averages
of the quantities of interest.

The wavelet decomposition using Daub4 is carried out in a pyramidal scheme such that
at level $j$ there is a total of $2^{j}$ coefficients: $a_{jk}(t)$ with
$k=1,\dots,2^j$ and $j=0,\dots,J$, except for the coarsest level $j=0$ where there
are $2$ coefficients, namely $a_{00}(t)$ and $a_{01}(t)$, instead of just one. To be
more precise in the notation, we define the wavelet contribution at level $j$ as
\begin{equation} v_j(x,t)=\sum_k a_{jk}(t)\Psi_{jk}(x),
\end{equation}
such that the field is a sum of all its wavelet contributions, $v(x,t)=\sum_j
v_j(x,t)$. Notice that as the wavelet level becomes coarser, i.e reducing $j$ by one,
the number of wavelet coefficients contributing to $v_j(x,t)$ is therefore reduced by
a factor of two and that the total number of wavelet coefficients $a_{jk}(t)$ equals
the number of points in the system, $N=2^{J+1}$.

\section{Wavelet Analysis: Results}
\subsection{Energy Distribution}
In the language of signal processing, the total energy of a signal
field is a conserved quantity.  For our case, the energy for the
chaotic field $v(x,t)$ can be defined as
\begin{equation}
E(t)=L^{-1}\langle v \mid v \rangle =L^{-1}\int_{0}^{L} v^{2}(x,t)dx
\end{equation}
and it can be written as the sum of the energies at each wavelet level, $E(t)=\sum_j
e_j(t)$ with $e_{j}(t)=L^{-1}\langle v_j \mid v_j \rangle $.  Fig. (1a) shows the
energy distribution per wavelet level for the Nikolaevskii model, averaged over
time.  One sees that the energy of the field is mainly concentrated at wavelet level
$j=5$. As a comparison, we also show the structure factor $S(q)$, which can be
thought of as the energy distribution in Fourier space, in Fig. (1b).  It is easy to
show for this model \cite{TV96} that the unstable modes (within linear stability
analysis) in Fourier space are located between $q_{1,2}=(1\pm
\sqrt{\epsilon})^{1/2}$, with the most unstable mode at $q_{m}\simeq 1.0$, consistent
with the results shown in Fig. 1(b).  In our case with $\epsilon=0.5$ it is
$q_{1}\simeq 0.54$ and $q_{2}\simeq 1.31$, so with $L=158.72$, the most energetic
mode should be at $n_{m}=q_{m}L/2\pi\simeq 25.26$ and the smallest and largest
unstable modes are at $n_1=q_{1}L/2\pi \simeq 13.67$ and $n_2= q_{2}/2\pi\simeq
33.00$ respectively.  Therefore there should be about $20$ complex or $40$ real
unstable or marginal modes concentrated in the neighborhood of the $50$th real mode.
Notice that wavelet levels $0-4$ contain a total of $32$ modes (corresponding to $32$
wavelet coefficients).  It is easy to understand, therefore, that the energy peak is
located at wavelet level $j=5$.  On the other hand, if one compares Figures (1a) and
(1b), we can also see that for small q there is no similarity between the two energy
distributions, implying that for small $q$ the overlap between different wavelet
levels is relatively strong.

\subsection{Temporal Behavior and Probability Distribution of Wavelet Coefficients}

Figure 2 shows the temporal behavior of some wavelet coeffients $a_{jk}(t)$. At each
level $j$ we show the intermediate wavelet coefficient with $k=2^{j-1}$. Notice that,
from the very early times, there is a
 chaotic behavior in the temporal evolution. In Fig. 3 we show the probability density

 functions (PDFs) for the wavelet
coefficients from level $j=1$ to $j=7$, averaged over time. Namely, for fixed j we
sum over all k for the  $a_{jk}$ coefficients to obtain a mean value; we repeat this
over all times sampled and plot these mean values to obtain the PDF.  The PDFs for
$j=0$ and $j=8$ are not shown here, since they are just Dirac-delta functions with
spikes at zero. The delta function-like behavior at the largest wavelet level $j=8$
simply means that there is a strong dissipation at large wavenumbers in the Fourier
spectrum (cf Wittenberg and Holmes \cite{WH99}). (In Fig. 3 we can already see a
similar behavior developing at the level $j=7$.)  The spike at $j=0$, however, only
implies that the coarsest part is not suitable for describing the field and its two
components compete with each other everywhere to yield a zero average value for the
amplitude. We have tried to fit these PDFs to a Gaussian distribution.  In Fig. 3 we
see that for $j=1, ..., 7$ the PDFS are essentially Gaussian. (We do not think the
small deviations from Gaussian are statistically significant.)

The wavelet decomposition of the STC in this model allows us to examine the chaotic
behavior of the system at different spatial scales.  In general, as can be seen
somewhat in Fig. 1a), the large $j$ levels correspond to large wavenumbers or small
spatial scales and the small $j$ levels correspond to small wavenumbers or large
spatial scales.  In this section we try to confirm this in more detail by showing the
structure factors which result from removing various wavelet levels.

In Fig.(4a-f) we show the structure factor for some of these "reduced level"
systems, with the full structure factor for the Nikolaevskii model  shown for
comparison. Fig. (4a) shows the result of removing wavelet level $j=8$, for
which the new structure factor is almost identical with the exact one. This
shows that removing the smallest spatial scale in the wavelets (or,
equivalently, the shortest wavelengths) leaves the system invariant. Fig. (4b)
shows the result of removing levels $j=7$ and $j=8$. In this case, there is a
small difference between the two structure factors for large wavenumbers,
indicating that the large $j$ levels of the wavelets only contribute to the
short-wavelength dynamics (i.e. the fast dissipation) of the model.  Fig. (4c)
shows the result of removing levels $j=6, 7$ and $8$.  In this case there is a
significant difference between the two structure factors near the peak.  This
is not surprising since $j=6$ is the second most energetic level in the model.
Fig. (4d) shows the results of removing levels $j=0,1$ and $2$.  The resulting
structure factors shows that these levels are responsible for the small peak in
the structure factor near wavenumber $q\approx 0$.  Fig. (4e) shows the role of
the most energetic levels in the model, i.e. $j=4,5$ and $6$.  We see that
these levels are responsible for the dominant part of the structure factor.
Furthermore, we also see that without level $j=3$, the small peak near
$q\approx 0$ becomes even smaller in comparison with Fig.(4d).  Finally, Fig.
(4f) shows the result of removing all levels with $j > 2$.  As one would
expect, one only has the peak near small $q$.  If one considers all of the
above results, one obtains a clear picture of the scale localization involved
in the wavelet decomposition.

We end this section with a discussion of the relationship between wavelet
decomposition and the renormalization group transformation. It is clear that removing
the large-$j$ wavelet levels is a procedure that essentially removes the
short-wavelength part of the structure factor in Fourier space.  In this sense it is
like a renormalization group transformation in which one integrates out the short
wavelength degrees of freedom to obtain a renormalized description of a system.  The
particular Daub4 algorithm we use here is a kind of decimation transformation in that
at every step removing a wavelet of the largest remaining level means that one has
only half as many spatial points as before the removal to describe the (chaotic)
field.  The other set of values of the field is "filtered out", with the value of the
field at a given new (remaining) space point being a kind of average of the field at
four of the neighboring points of the original set. Other types of wavelet
decompositions correspond to different kinds of transformations.

\subsection{Lyapunov Spectrum}

We conclude by examining the effect of removing wavelet levels on the Lyapunov
spectrum of the system.  This provides us with the most detailed understanding
of the contribution of various levels to the chaotic dynamics of the system.
Fig. (5a) shows the Lyapunov spectrum obtained after removing the $j=8$ level,
while Fig. (5b) shows the Lyapunov spectrum obtained after removing $j=7$ and
$j=8$. The Lyapunov spectrum for the Nikolaevskii model is also shown for
comparison.  In Table 1 we show the Lyapunov dimension $D$ and the
Kolmogorov-Sinai entropy $H$ for these different cases. It seems that to a very
good approximation the chaotic behavior of the system does not depend on the
$j=8$ level. Thus a simplified statistical description of the Nikolaevskii
equation can be obtained by excluding the $j=8$ level, without altering the
chaotic dynamics for the system.  However, it is also clear from inspection of
Table 1 that removing any additional levels significantly alters the behavior.
Thus all these levels play an important role in the spatiotemporal chaotic
dynamics of the system.

\section{Conclusion}

We have shown that the dynamics of the one dimensional Nikolaevskii equation in the
spatiotemporally chaotic regime may be understood in terms of localized dynamics in
both space and scale (wavenumber). Specifically, a projection onto a particular
Daubechies basis (Daub4) yields a good separation of scales, as shown by an
examination of the contribution of different wavelet levels to the power spectrum. At
most scales, including the most energetic ones, we find essentially Gaussian
dynamics.  Perhaps most importantly, we also found that removal of certain wavelet
modes can be made without altering the chaotic dynamics of the system as described by
the Lyapunov spectrum.

Many different length scales have been proposed for the description of spatiotemporal
chaos \cite{CH93}. These include the usual correlation length $\xi_{2}$ for the (two
point) order parameter and the dimension correlation length $\xi_{\delta}$, obtained
from the Lyapunov dimension D(L) as $\xi_{\delta}=lim_{L\rightarrow \infty}
L^d/D(L)$. Here $d$ denotes the space dimensionality and $L$ the linear dimension of
the system.  We have studied both of these lengths as a function of the control
parameter $\epsilon$ for this model. The results of these calculations, as well as
for other quantities characterizing the spatiotemporal chaos will be presented
elsewhere \cite{TXGT00}. Since it is difficult to calculate the Lyapunov dimension for
high-dimensional systems, Zoldi and Greenside have proposed using a Karhunen-Loeve
dimension $D_{KLD}$, defined by the number of eigenmodes in a proper orthogonal
decomposition necessary to capture a given fraction $f$ of the total energy.  From
this dimension one can define a Karhunen-Loeve correlation length $\xi_{KLD}$.
However, for a translationally invariant system, such as the Nikolaevskii equation
with periodic boundary conditions, the Karhunen-Loeve eigenmodes are Fourier modes.
Thus the Karhunen-Loeve dimension length for any $f$ can be computed directly from
the power spectrum $S(q)$ and thus contains no more dynamical information than
$\xi_{2}$. As Wittenberg and Holmes have pointed out \cite{WH99} all these lengths
(with the possible exception of the dimension correlation length $\xi_{\delta}$) are
measures only of spatial disorder and thus yield no information about the temporally
chaotic dynamics responsible for the disorder.  As a consequence, Wittenberg and
Holmes introduced another length scale, a so-called dynamical interaction length.
This is a length scale $l_{c}$ such that if one deletes interactions for length
scales greater than $l_{c}$ in the wavelet Galerkin projection of the model equation
of interest, one alters the chaotic dynamics of the model (i.e. changes the Lyapunov
spectrum). Although this is in principle a very interesting length scale to study, it
involves a numerical calculation which for our model is computationally expensive and
beyond our current resources.  As a consequence, we leave this important issue for
future investigation.

Finally, as mentioned in the introduction,  Wittenberg and Holmes \cite{WH99}
performed a wavelet decomposition of the Kuramoto-Sivashinsky model, using an
orthogonal spline basis. They found that the probability distribution functions
for the most energetic modes were non-Gaussian. This differs from our results
for the Nikolaevskii model, in which the distribution functions for the most
energetic modes are Gaussian (to a very good approximation and excluding the
peak near zero). To see whether this difference is significant or simply
results from a different choice of basis functions in the two studies, we have
calculated the PDFs for the KS model using the Daub4 basis.  We found that with
this choice of basis the PDFs for the most energetic modes are in fact
Gaussian. We find non-Gaussian behavior only at the small-$j$ levels.  Thus it
would seem that any non-Gaussian behavior of the wavelet coefficients depends
on the basis chosen for the wavelet decomposition.

{\bf Acknowledgments}  This work was supported by NSF grant DRM9813409 and DGES
(Spain) projects PB97-0141-C02-01 and BMF2000-1108. One of us (GX) would like
to acknowledge the support of a grant from the China Scholarship Council.\\
{*}Permanent address: Department of Applied Physics, Sichuan University, Chengdu 610065 China.

\newpage

\begin{tabular}{|c|c|c|} \hline
Wavelet levels  & Lyapunov dimension & KS entropy \\ \hline including j=0-8 & 52.31 &
5.27\\ \hline including j=0-7 & 51.89 & 5.15 \\ \hline including j=0-6 & 51.40 & 4.81
\\ \hline including j=0-5 & 57.42 & 6.82 \\ \hline including j=0-4 & 79.64 & 3.28 \\
\hline including j=4-8 & 57.10 & 8.48 \\ \hline including j=4-6 & 57.00 & 7.95 \\
\hline including j=0-2,4-8 & 55.13 & 8.01 \\ \hline
\end{tabular}
\vfill
\eject
\newpage
\makebox{\epsfbox{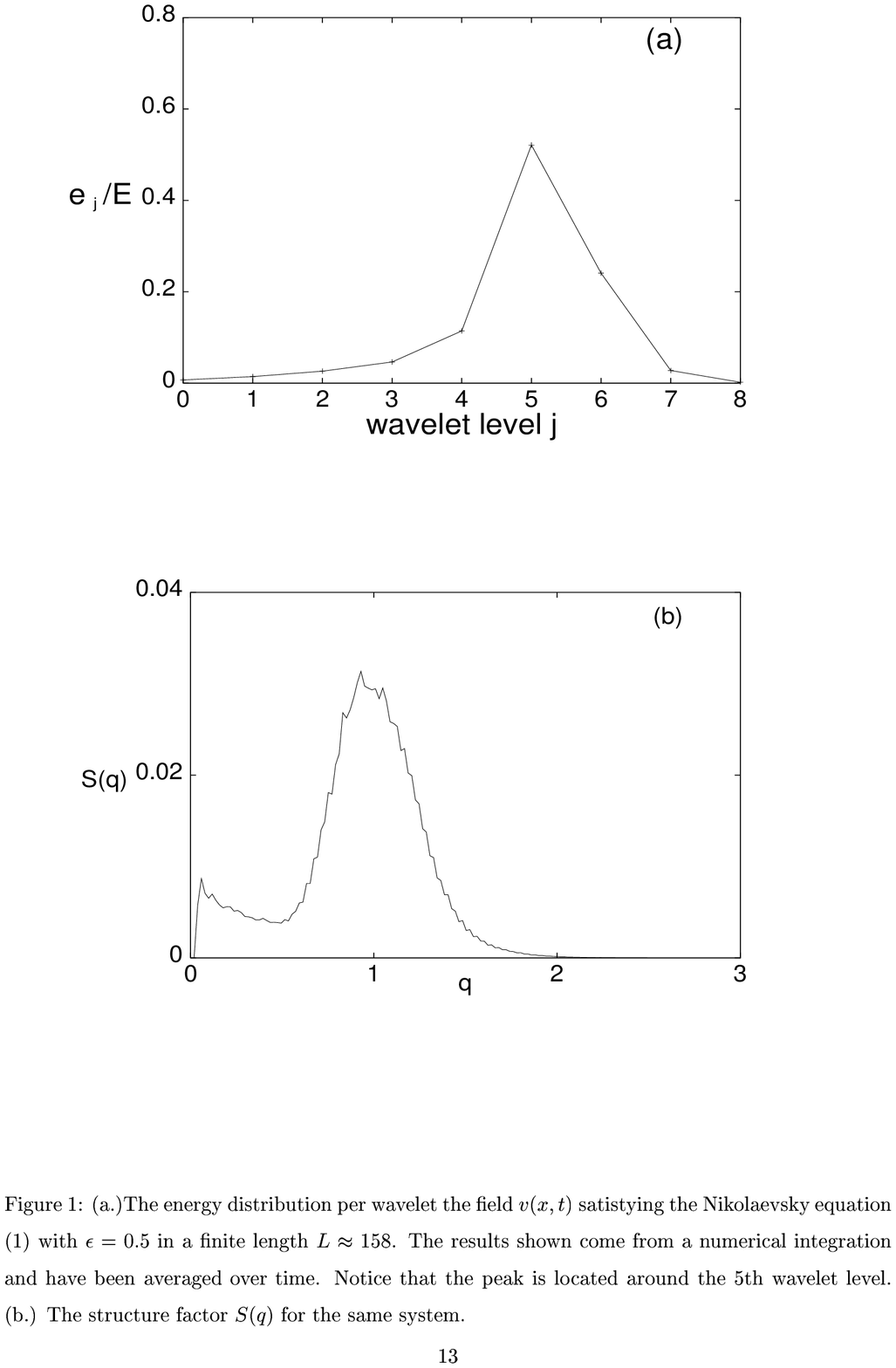}}
\vfill
\newpage

\makebox{\epsfbox{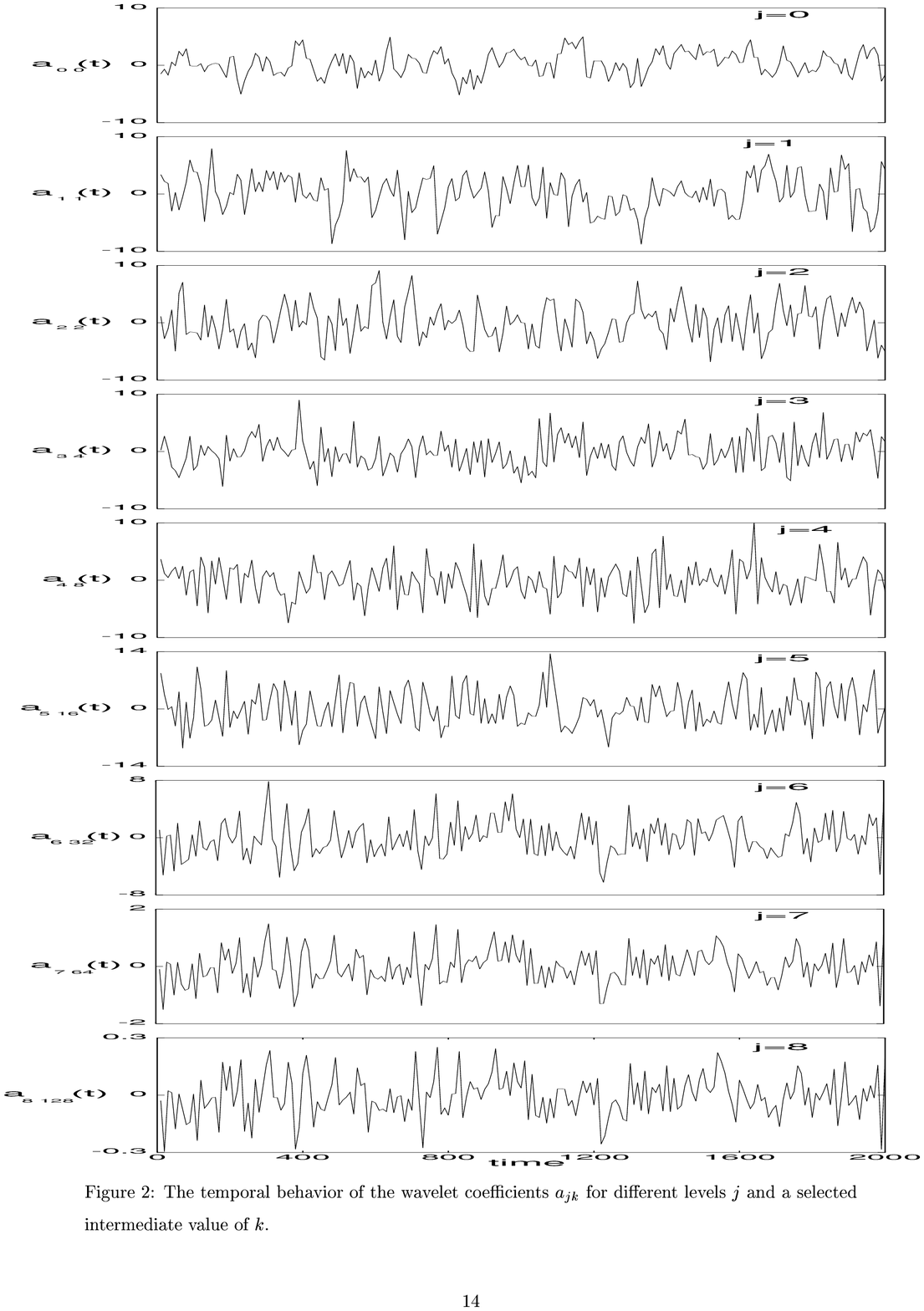}}
\newpage

\makebox{\epsfbox{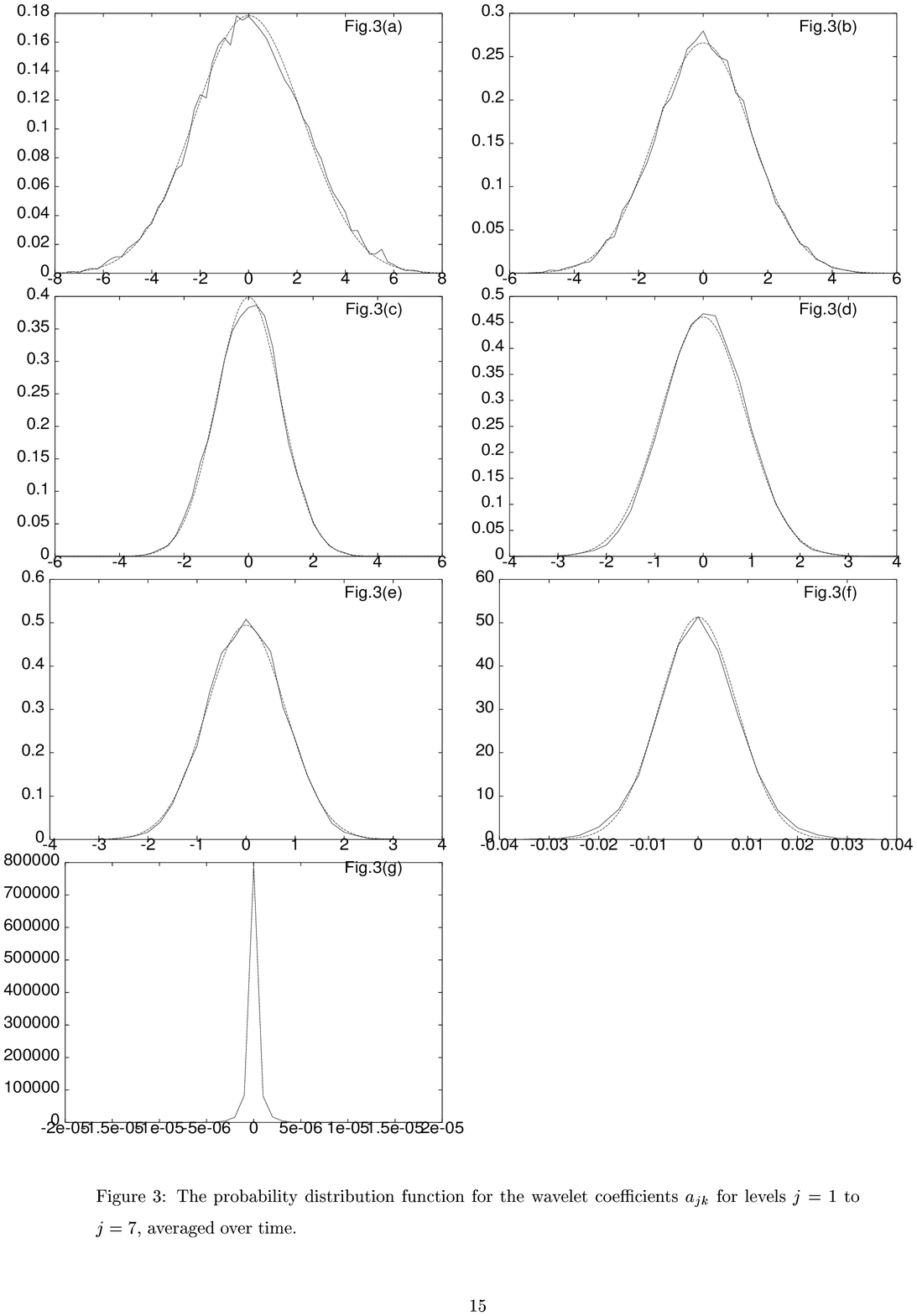}}
\newpage

\makebox{\epsfbox{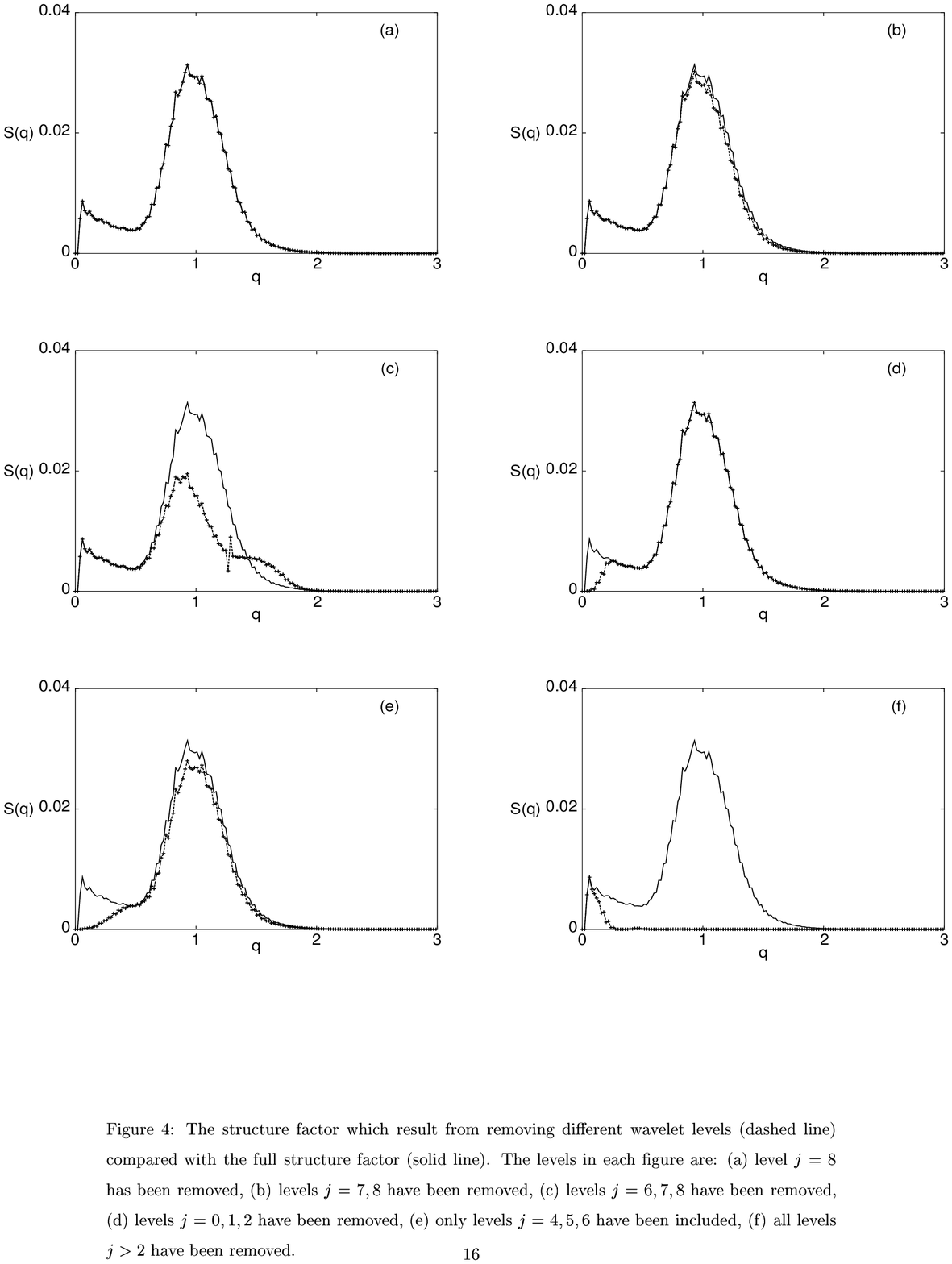}}
\newpage

\makebox{\epsfbox{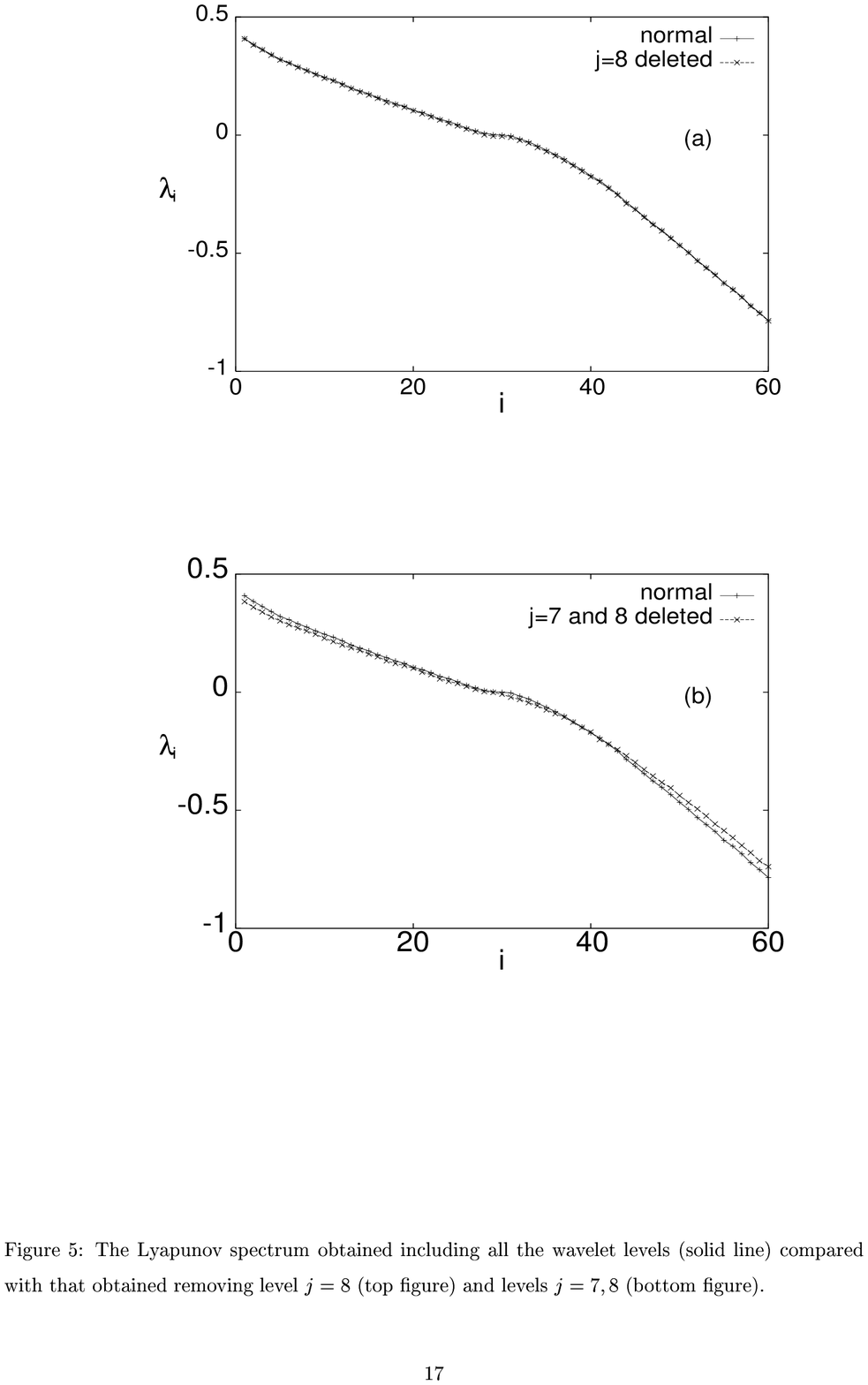}}

\end{document}